\documentclass[runningheads]{llncs}
\usepackage{graphicx}
\usepackage{subfig}
\usepackage{amsmath}
\usepackage{hhline}
\usepackage{xkeyval,xcolor}
\makeatletter
\newlength{\sfp@hseplen}\newlength{\sfp@vseplen}
\define@cmdkey{subfigpos}[sfp@]{pos}[ul]{}
\define@cmdkey{subfigpos}[sfp@]{font}[\small]{}
\define@cmdkey{subfigpos}[sfp@]{vsep}[5pt]{\setlength{\sfp@vseplen}{\sfp@vsep}}
\define@cmdkey{subfigpos}[sfp@]{hsep}[5pt]{\setlength{\sfp@hseplen}{\sfp@hsep}}
\newcommand{\subfigimg}[3][,]{%
  \setkeys{Gin,subfigpos}{pos,font,vsep,hsep,#1}
  \setbox1=\hbox{\includegraphics{#3}}
  \ifnum\pdfstrcmp{\sfp@pos}{ul}=0
    \leavevmode\rlap{\usebox1}
    \rlap{\hspace*{\sfp@hsep}\raisebox{\dimexpr\ht1-\sfp@vsep}{\sfp@font{#2}}}
    \phantom{\usebox1}
  \else\ifnum\pdfstrcmp{\sfp@pos}{ur}=0
    \leavevmode\usebox1
    \llap{\raisebox{\dimexpr\ht1-\sfp@vsep}{\sfp@font{#2}}\hspace*{\sfp@hsep}}
  \else\ifnum\pdfstrcmp{\sfp@pos}{lr}=0
    \leavevmode\usebox1
    \llap{\raisebox{\sfp@vsep}{\sfp@font{#2}}\hspace*{\sfp@hsep}}
  \else
    \leavevmode\rlap{\usebox1}
    \rlap{\hspace*{\sfp@hseplen}\raisebox{\sfp@vsep}{\sfp@font{#2}}}
    \phantom{\usebox1}
  \fi\fi\fi
}
\makeatother

\begin{document}
\title{Unsupervised Domain Adaptation for\\Small Bowel Segmentation using\\Disentangled Representation}
%
\titlerunning{Unsupervised Domain Adaptation for Small Bowel Segmentation}
%
\author{Seung Yeon Shin \and
Sungwon Lee \and
Ronald M. Summers}
%
\authorrunning{S.Y. Shin et al.}
%
\institute{Imaging Biomarkers and Computer-Aided Diagnosis Laboratory, Radiology and Imaging Sciences, Clinical Center, National Institutes of Health, Bethesda, MD, USA\\
\email{\{seungyeon.shin,rms\}@nih.gov}}
\maketitle              
%
\begin{abstract}
We present a novel unsupervised domain adaptation method for small bowel segmentation based on feature disentanglement. To make the domain adaptation more controllable, we disentangle intensity and non-intensity features within a unique two-stream auto-encoding architecture, and selectively adapt the non-intensity features that are believed to be more transferable across domains. The segmentation prediction is performed by aggregating the disentangled features. We evaluated our method using intravenous contrast-enhanced abdominal CT scans with and without oral contrast, which are used as source and target domains, respectively. The proposed method showed clear improvements in terms of three different metrics compared to other domain adaptation methods that are without the feature disentanglement. The method brings small bowel segmentation closer to clinical application.

\end{abstract}
\section{Introduction}\label{sec:intro}
The small bowel is a part of the gastrointestinal tract between the stomach and the large bowel. It ranges from 20 to 30 feet long but is highly convoluted so that it can fit into the abdominal cavity~\cite{cc19}. Apart from its simple tubular structure, it has variable configuration while surrounded by visceral fat and other organs including the large bowel. Also, the appearance may differ locally according to the internal material, e.g., gas and fluid.

Computed tomography (CT) has been considered the first-line imaging modality for the evaluation of small bowel diseases since it is fast and non-invasive compared to other imaging tests such as endoscopy while providing essential diagnostic information~\cite{murphy14}. Despite the clinical benefit, the interpretation itself is laborious and time-consuming. Automatic segmentation of the small bowel could expedite the interpretation. Specifically, it may help precise localization of diseases, such as inflammatory bowel disease and carcinoid tumors, and preoperative planning by better visualization.

\begin{figure}[t]
	\centering
	\begin{minipage}{1\textwidth}
        \subfigimg[width=0.25\textwidth,pos=ll, font=\color{white}]{A}{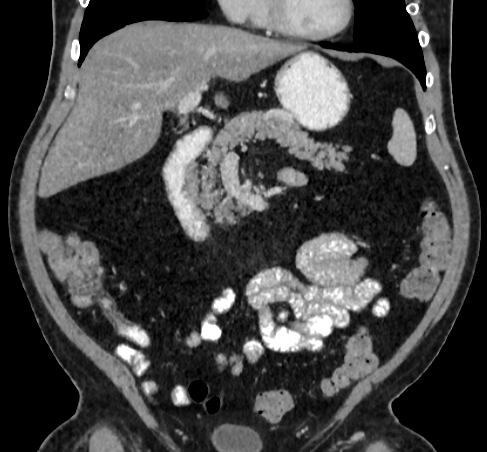}
        \hspace{-0.2cm}
        \subfigimg[width=0.25\textwidth,pos=ll, font=\color{white}]{B}{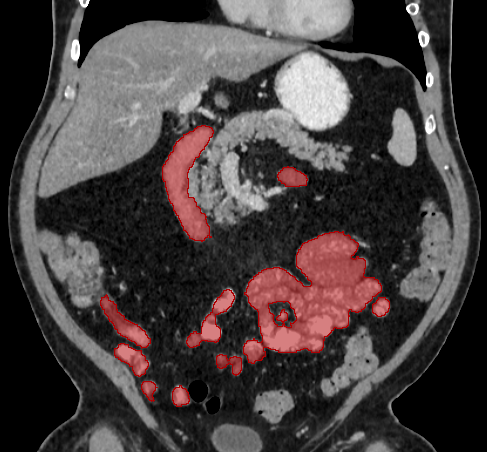}
        \hspace{-0.2cm}
        \subfigimg[width=0.25\textwidth,pos=ll, font=\color{white}]{C}{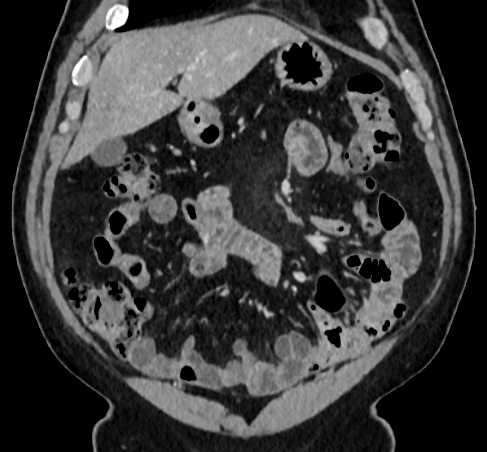}
        \hspace{-0.2cm}
        \subfigimg[width=0.25\textwidth,pos=ll, font=\color{white}]{D}{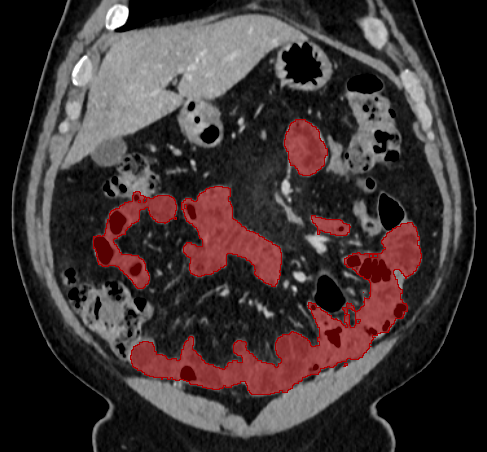}
    \end{minipage}
	\caption{Example CT scans (A) with and (C) without oral contrast. The respective ground-truth segmentation of the small bowel is shown as red in (B) and (D).}
	\label{fig:ex_img}
\end{figure}

Over the years, there have been attempts to develop automatic methods for small bowel segmentation, especially using deep learning. The small bowel was included in segmenting multiple organs-at-risk for radiotherapy treatment planning of affected tissues, such as pancreatic and cervical cancers, in CT scans~\cite{ju20,liu20,sartor20}. Although the results obtained for the small bowel are reasonable, some of their data included only the part of the small bowel that is closest to the target area, which needed to be dose-evaluated~\cite{ju20,liu20}. In \cite{sartor20}, the rough bowel location was detected instead of performing pixel-accurate small bowel segmentation. There have been only a few previous works dedicated solely to automatic small bowel segmentation~\cite{oda20,shin20,zhang13}. While the specific anatomic relationship between the mesenteric vasculature and the small bowel is used to guide the small bowel segmentation in \cite{zhang13}, a cylindrical shape constraint is applied during training of the small bowel segmenter in \cite{shin20}. Although each of the works showed reasonable performance for particular datasets, their generalizability across different datasets was not evaluated.
A CT scan is acquired using a specific imaging protocol depending on the purpose of the investigation, which includes the use of different contrast media and scan timing. Thus, the appearance of the small bowel may be different across datasets as exemplified in Figure~\ref{fig:ex_img}.
It is observed in our experiment that, when trained on one dataset, the model does not generalize well to another dataset due to the domain shift (section~\ref{sec:results}).

Domain adaptation is a task to address the domain shift problem and has been gaining attention in various fields. In this work, an unsupervised domain adaptation scenario, where ground-truth (GT) labels are not available for the target domain, is considered. Recent unsupervised domain adaptation methods can be categorized into two groups according to which specific space is to be aligned between different domains: 1) input data space~\cite{yang19,zhu17} and 2) output space~\cite{chen17,tsai18,wang19}. In the first group works, image-to-image translation~\cite{zhu17} is used to translate images from target domains to a source domain. Then, the translated images can be tested using a source domain model. Cross-modality liver segmentation was performed by translating between CT and MRI images in \cite{jiang20,yang19}. On the other hand, adversarial learning is used to encourage the output prediction of the target domain to be similar to the source ones in the second group works. This adaptation can be applied in the feature level~\cite{chen17}, multiple output levels~\cite{tsai18}, or multiple kinds of outputs~\cite{wang19}. Our method falls into the second group. Domain adaptation is even more important for small bowel segmentation since it is very hard to achieve GT labels for multiple datasets due to the high difficulty of the labeling. In practice, relatively small numbers (ten or less) of annotated CT scans were used in recent works~\cite{oda20,shin20}. It would be beneficial if it is possible to adapt a network to the target dataset without the use of labels.

In this paper, we present a novel unsupervised domain adaptation method for small bowel segmentation, which is based on feature disentanglement. Although the absolute intensity values in CT scans (Hounsfield units) carry important information on specific substances of the human body, thus could provide a clue of being specific tissues and organs, they may be variable according to the imaging protocol. Figure~\ref{fig:ex_img} shows example CT scans that were acquired with and without oral contrast administration. The absolute intensity values are no longer a strong clue for the small bowel when we train and test across the datasets. Non-intensity features like texture and shape may be more useful.
For example, local textures of the valvulae conniventes, which are circular folds on the inner surface of the small bowel, are more recurrent across the datasets.

Disentangling feature representations into desired factors provides not only an understanding of a deep network, but also more controllability on it. However, to achieve it in an unsupervised manner, either some prior assumptions or proper modification on the network architecture is required~\cite{berga20,shu18}. For example, when decomposing a face image into appearance and deformation components, a smoothness constraint is applied on the inferred deformation in \cite{shu18}. In our method, feature disentanglement is performed using a unique auto-encoding architecture paired with augmented input images, without any prior assumption.

From the observation that the non-intensity features would be more transferable across the datasets than the intensity features, we first disentangle them within the proposed auto-encoding architecture. Then, only the non-intensity features are guided to be domain-invariant using adversarial learning~\cite{goodfellow14}. Finally, segmentation prediction is performed by aggregating the disentangled features. To the best of our knowledge, this is the first work to develop an unsupervised domain adaptation method for small bowel segmentation. The proposed adaptation method based on the feature disentanglement further increased the adaptability of the segmenter, resulting in clear improvements for all evaluation metrics compared to the alternative methods.

\section{Method}\label{sec:method}

\subsection{Dataset}

Our dataset consists of intravenous contrast-enhanced abdominal CT scans which were done during the portal venous phase. It is composed of two subsets depending on whether or not oral contrast was used. The first subset includes 20 scans that were acquired with oral administration of Gastrografin, which is used as the source domain dataset. Meanwhile, the second subset, which includes 70 scans acquired without any oral contrast, is used as the target domain dataset. We resampled all the volumes to have isotropic voxels of $2mm^{3}$.
The images were cropped manually to include from the diaphragm through the pelvis.

GT labels were achieved using ``Segment Editor" module in 3DSlicer~\cite{fedorov12} by a radiologist with 12 years of experience. The GT segmentation includes the duodenum, jejunum, and ileum while not including any mesenteric fat, vessels, colon, and abdominal wall. We note that this annotation took several hours for each scan. We acquired GT segmentations for all 20 scans of the source domain dataset, and for 10 scans of the target domain dataset. All the annotated 10 scans are used as the test set, and the remaining 60 scans are used as unsupervised training samples.
We note that this number of GTs is bigger than that of the previous works~\cite{oda20,shin20}, which was ten or less.

\subsection{Unsupervised Disentangling of Intensity and Non-Intensity Representations}\label{subsec:disen}

\begin{figure}[t]
	\centering
	\includegraphics[width = 1\textwidth]{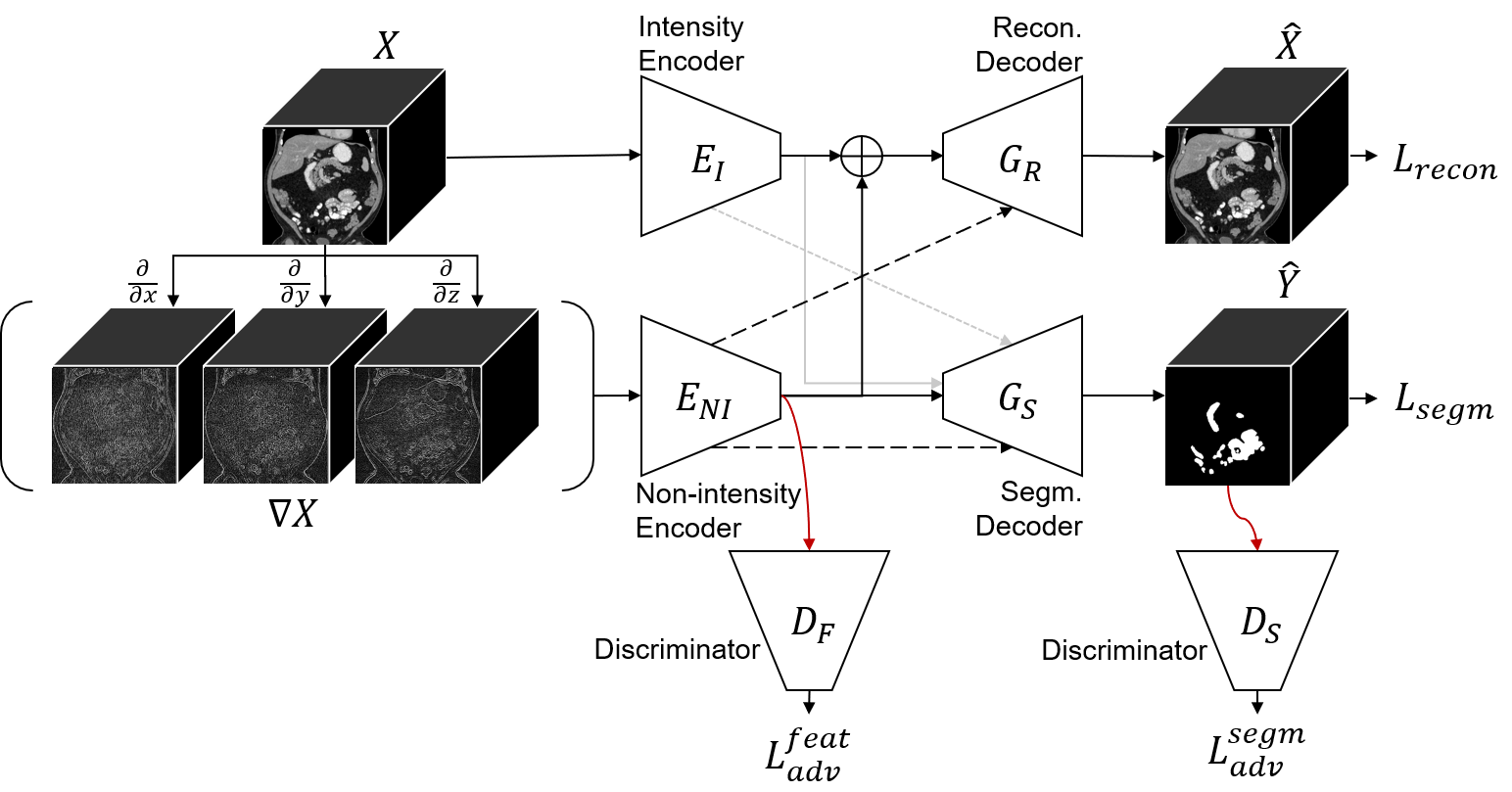}
	\caption{Network architecture for the proposed method. The network is composed of two encoders $E_I$, $E_{NI}$, and two decoders $G_R$, $G_S$. The two encoders separately extract the intensity and non-intensity features through the specific network design and augmented input. The two decoders reconstruct the input volume and predict the segmentation, respectively.
	$\nabla X$ represents the gradient images of $X$, and the concatenation of the gradient along each axis is fed into the non-intensity encoder. The dotted lines represent skip connections, and $\oplus$ means feature concatenation. The gray lines are optional connections, which can be disconnected in some mode. The two discriminator networks $D_F$, $D_S$, and four loss functions $L_{recon}$, $L_{segm}$, $L_{adv}^{feat}$, $L_{adv}^{segm}$ involved in training the proposed network are also shown. Refer to the text for the details.}
	\label{fig:network}
\end{figure}

Figure~\ref{fig:network} shows the proposed network composed of a sub-network for feature disentanglement and an additional decoder for segmentation prediction based on the disentangled features. The sub-network for feature disentanglement has a two-stream auto-encoding architecture, where the intensity and non-intensity features are first extracted through the separate encoders $E_I$ and $E_{NI}$, respectively, and then combined in the decoder $G_R$ to reconstruct the input image.

All convolution and pooling layers in the intensity encoder have 1x1x1 kernels to constrain it to see each voxel independently and learn only the intensity information. While the intensity encoder takes the original image $X$ as input, the non-intensity encoder uses as input the gradient images of $X$, $\nabla X = \{\frac{\partial X}{\partial x}, \frac{\partial X}{\partial y}, \frac{\partial X}{\partial z}\}$ as shown in Figure~\ref{fig:network}. As a simple operation, the gradient still maintains the non-intensity information like texture and shape by keeping the relative values of neighboring voxels while losing the absolute values in a CT scan. Thus, the non-intensity encoder is guided to learn the non-intensity features.

The extracted features from both encoders are finally combined to reconstruct the original image in the reconstruction decoder. To prevent this reconstruction from being done solely from the intensity feature, the following are applied: 1) While the lower level features of the non-intensity encoder are used by skip connections, only the innermost features from the intensity encoder are used for reconstruction. 2) Dropout is applied to the intensity features before concatenation with the non-intensity features. We applied the per-element dropout based on empirical evaluation. The loss function for reconstruction is as follows:

\begin{equation}
    \label{eq:loss_rec}
	L_{recon}(X) = \frac{1}{|X|}\sum_{i}|x_i-\hat{x}_i|,
\end{equation}
where $\hat{X}=\{\hat{x}_i\}_{i=1}^{h \times w \times l}$ is the reconstruction from the input $X$ of size $h \times w \times l$.

\subsection{Unsupervised Domain Adaptation using Disentangled Representations}

The key idea of this work is to perform the domain adaptation on the disentangled representation, but it is also done in the output level.

\subsubsection{Feature Level Adaptation}

Adversarial learning is applied to the disentangled non-intensity feature to make it domain-invariant while leaving the intensity feature as it is. The involved fully convolutional discriminator $D_F$ takes the innermost features $E_{NI}(\nabla X)$ as input, and outputs $P=D_F(E_{NI}(\nabla X))=\{p_j\}_{j=1}^{h' \times w' \times l'}$, where the value $p_j$ represents the probability that the image $X$ is drawn from the source domain for the corresponding position $j$ in the feature map. The cross-entropy loss $L_{disc}^{feat}$ for training $D_F$ is defined as:

\begin{equation}
    \label{eq:loss_disc_feat}
	L_{disc}^{feat}(X) = -\frac{1}{|P|}\sum_{j}\Big((1-z)\log p_{j} + z\log(1-p_{j})\Big),
\end{equation}
where $z=0$ or $1$ for the source and target domain images, respectively.

While training the discriminator $D_F$ using images from both domains, an adversarial loss is computed for target domain images $X^{tar}$. Given the discriminator output $P^{tar}=\{p^{tar}_j\}_{j=1}^{h' \times w' \times l'}$, the adversarial loss is defined as:

\begin{equation}
    \label{eq:loss_adv_feat}
	L_{adv}^{feat}(X^{tar}) = -\frac{1}{|P^{tar}|}\sum_{j}\log p^{tar}_{j}.
\end{equation}
The non-intensity encoder $E_{NI}$ is encouraged to learn domain-invariant features in order to fool the discriminator $D_F$ during training. 

\subsubsection{Output Level Adaptation}\label{subsubsec:out_level_adapt}

The segmentation prediction is performed in an additional decoder $G_S$ by aggregating the disentangled features.
Being less transferable across domains, the intensity features are optionally used for segmentation prediction, which is implemented by optional connections as shown in Figure~\ref{fig:network}.
The effect of this will be evaluated in section~\ref{sec:results}. We used the generalized Dice loss~\cite{sudre17} for $L_{segm}$ to train the segmentation decoder. The calculation of this supervised loss is possible only for source domain images, meaning that the segmentation decoder $G_S$ is fitted by only source domain images.

To adapt the segmentation decoder, another adversarial learning is applied to the segmentation output $\hat{Y}$. The loss $L_{disc}^{segm}$ for training the output level discriminator $D_S$, and the adversarial loss $L_{adv}^{segm}$ are defined similarly to ones for feature level adaptation but for the segmentation prediction $\hat{Y}$. This encourages the output prediction of the target domain to be similar to the source ones.

\subsubsection{Objective Function for Domain Adaptation}

Finally, the overall loss function for training the proposed network is:

\begin{equation}
    \label{eq:loss_tatal}
    \begin{aligned}
	L_(X^{src},X^{tar}) = & L_{segm}(X^{src}) + \lambda_{recon}L_{recon}(X^{src},X^{tar}) + \\  & \lambda_{adv}^{feat}L_{adv}^{feat}(X^{tar}) + \lambda_{adv}^{segm}L_{adv}^{segm}(X^{tar}),
	\end{aligned}
\end{equation}
where $\lambda_{recon}$, $\lambda_{adv}^{feat}$, and $\lambda_{adv}^{segm}$ are the weight for each loss.

\subsection{Evaluation Details}

In the proposed network, the structure of each encoder and decoder is based on that of the 3D U-Net~\cite{cicek16}, but has a smaller number of channels, which are \{32, 64, 128, 256\}. All convolution layers have 3x3x3 kernels excepting ones in the intensity encoder and the final inference layer in the decoders, which have 1x1x1 kernels. Group normalization~\cite{wu18} is used between each convolution layer and non-linearity function. The feature level discriminator network $D_F$ consists of three convolution layers with 3×3×3 kernels and one final convolution layer with 1×1×1 kernels, where the numbers of channels are {32, 64, 128, 1}, respectively. A pooling layer is added between each convolution layer. The output level discriminator $D_S$ is with a similar structure, but has one more convolution layer, where the numbers of channels are {32, 64, 128, 256, 1}.

We implemented all the networks, including ones for the comparable methods, using PyTorch 1.2. For training, we used AdamW optimizers~\cite{loshchilov19} and a weight decay of $5 \times 10^{-4}$. Based on the grid search, the learning rates of $10^{-4}$ and $10^{-6}$ were used for the encoders and decoders, and for the discriminators, respectively. We used $0.2$, $10^{-6}$, $10^{-4}$ for $\lambda_{recon}$, $\lambda_{adv}^{feat}$, $\lambda_{adv}^{segm}$, respectively. Dropout with a probability of 0.3 is applied to the intensity features before concatenation with the non-intensity features as described in section~\ref{subsec:disen}.

We used a NVIDIA Tesla V100 32GB GPU to conduct experiments. To fit in the memory, sub-volumes of size 112x112x112 sampled from the original volume are used during training. The mini-batch size was set as $1$. We applied image rotation, elastic transformation, random global intensity adjustment for data augmentation. Dice coefficient, 95\% Hausdorff distance (HD95), and average symmetric surface distance (ASD) are used as evaluation metrics. Also, paired t-tests are conducted to show the statistical significance of the proposed method.

\section{Results}\label{sec:results}

\subsection{Quantitative Evaluation}

Table~\ref{tab:quan_res} shows quantitative results of the proposed method and other unsupervised domain adaptation methods on the target domain dataset. We first provide the segmentation performance without any domain adaptation, which are: 1) performing a 5-fold cross validation within the target domain test set (tar 5-fold CV), 2) applying a model trained from the source domain to the target domain images (w/o DA), and 3) the same with 2) but using the gradient images instead of the original image as input. The use of the gradient images slightly improved the generalizability by preventing the network from seeing absolute intensity values in CT scans. The cross validation within the target domain test set was expected to perform better than other unsupervised domain adaptation methods since it is fully supervised. However, it is worse than most of the presented domain adaptation methods. When only a small amount of labels are accessible in the target domain, rather than attempting to train a `strictly' supervised model, it may be better to adapt a model from a different domain to the target domain using relatively abundant unlabeled target domain images.

For the other domain adaptation methods, while the same network architecture based on the 3D U-Net~\cite{cicek16} is used, adaptation is performed in a different way, i.e., either in the feature level or in the output level, or both. The effect of using the gradient images is evaluated again for the method that is with both feature and output level adaptation (feat. \& out. level DA w/ grad.), showing a worse result compared to using the original image. It corresponds to the proposed network without the intensity encoder and reconstruction decoder, which play an important role for feature disentanglement. The proposed method, where domain adaptation is performed on the disentangled representation, shows the best performance in terms of all metrics. We note that, when the segmentation prediction is performed without using the intensity features, our method has the same model complexity with the others since the intensity encoder and reconstruction decoder are no more involved in inference time. The p-values computed by conducting paired t-tests between the proposed method and the others with the Dice coefficients show the statistical significance of the proposed method.

As mentioned, segmentation prediction is performed by aggregating the disentangled features in our network. The variant using the intensity features as well as the non-intensity features (Ours + int. feat.), through the optional connection in Figure~\ref{fig:network}, does not improve the performance.
It implies that the disentangled non-intensity features are enough to perform the cross-domain small bowel segmentation.
To validate the need of feature and output level adaptation in the proposed method, the variants without either of them are also evaluated. While both contribute to achieving the best performance, the feature level adaptation that is applied directly on the disentangled non-intensity features is more important to guide the network to operate as we expect.

\begin{table}[t]
\centering
\caption{Comparison with other methods on the target domain test set. The first three are without any domain adaptation (DA). The remaining are different domain adaptation methods including the proposed method and its variant. Refer to the text for the explanation on each method. For every metric, the mean and standard deviation are presented. P-values are computed by conducting paired t-tests between the proposed method and the others with the Dice coefficients.}\label{tab:quan_res}
\begin{scriptsize}
\begin{tabular}{c|c|c|c|c}
Method & Dice & HD95 (mm) & ASD (mm) & p-value \\
\hline
tar 5-fold CV & 0.809 $\pm$ 0.081 & 13.024 $\pm$ 5.829 & 2.998 $\pm$ 1.315 & 0.004\\
\hline
w/o DA & 0.725 $\pm$ 0.141 & 15.944 $\pm$ 6.707 & 3.622 $\pm$ 1.657 & $1.380\times10^{-5}$\\
\hline
gradient input & 0.749 $\pm$ 0.125 & 17.405 $\pm$ 10.093 & 4.178 $\pm$ 2.077 & $2.725\times10^{-6}$\\
\hhline{=====}
feature level DA~\cite{chen17} & 0.807 $\pm$ 0.084 & 13.467 $\pm$ 6.053 & 3.040 $\pm$ 1.094 & $1.006\times10^{-5}$\\
\hline
output level DA~\cite{tsai18} & 0.813 $\pm$ 0.106 & 15.355 $\pm$ 8.063 & 3.243 $\pm$ 1.864 & 0.016\\
\hline
multi output level DA~\cite{tsai18} & 0.816 $\pm$ 0.098 & 14.599 $\pm$ 11.562 & 3.121 $\pm$ 2.048 & $9.270\times10^{-5}$\\
\hline
feat. \& out. level DA & 0.814 $\pm$ 0.100 & 12.487 $\pm$ 6.547 & 2.892 $\pm$ 1.480 & 0.006\\
\hline
feat. \& out. level DA w/ grad. & 0.811 $\pm$ 0.063 & 17.948 $\pm$ 9.994 & 3.527 $\pm$ 1.494 & 0.002\\
\hhline{=====}
Ours & \textbf{0.837} $\pm$ 0.084 & \textbf{10.290} $\pm$ 8.057 & \textbf{2.388} $\pm$ 1.408 & -\\
\hline
Ours + int. feat. & 0.837 $\pm$ 0.085 & 10.559 $\pm$ 6.702 & 2.536 $\pm$ 1.207 & 0.860 \\
\hline
Ours w/o output level DA & 0.829 $\pm$ 0.073 & 14.024 $\pm$ 7.830 & 2.940 $\pm$ 1.059 & 0.137 \\
\hline
Ours w/o feature level DA & 0.820 $\pm$ 0.083 & 14.745 $\pm$ 8.760 & 3.085 $\pm$ 1.375 & 0.006 \\
\end{tabular}
\end{scriptsize}
\end{table}

\subsection{Qualitative Evaluation}

\begin{figure}[t]
	\centering
	\begin{minipage}{1\textwidth}
        \subfigimg[width=0.33\textwidth,pos=ll, font=\color{black}]{A}{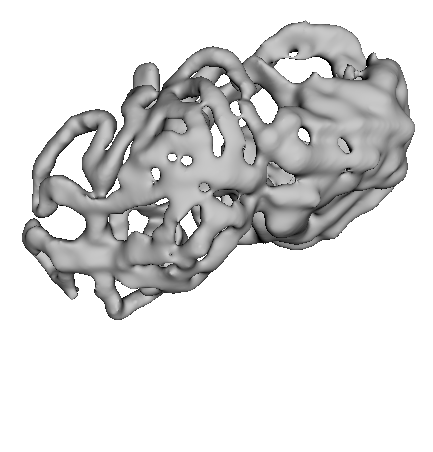}
        \hspace{-0.15cm}
        \subfigimg[width=0.33\textwidth,pos=ll, font=\color{black}]{B}{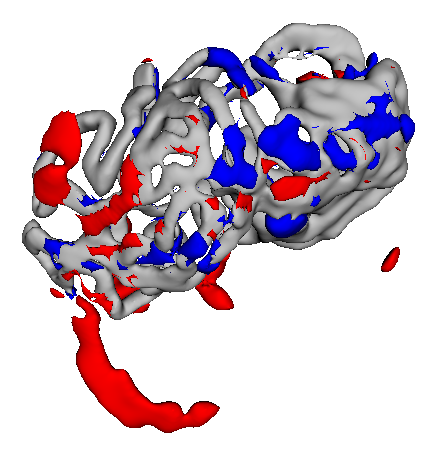}
        \hspace{-0.15cm}
        \subfigimg[width=0.33\textwidth,pos=ll, font=\color{black}]{C}{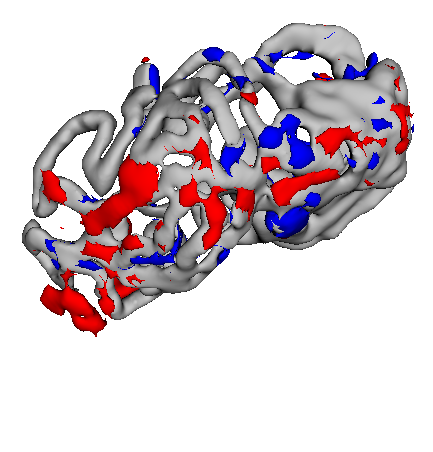}
    \end{minipage}
	\caption{Example segmentation results in 3D. (A) Ground-truth segmentation. (B) Result corresponding to `feat. \& out. level DA' in Table~\ref{tab:quan_res}. (C) Result of the proposed method. In (B) and (C), each result is compared with the ground-truth, and false positives and false negatives are marked as red and blue, respectively.}
	\label{fig:res_3d}
\end{figure}

Figure~\ref{fig:res_3d} shows example segmentation results in 3D. The result corresponding to ‘feat. \& out. level DA’ in Table~\ref{tab:quan_res} is compared to ours. We note that the only difference between them is whether the feature disentanglement is involved for the domain adaptation, thus could show its effectiveness. Fewer errors are observed for the proposed method. We believe this is because the proposed method explicitly concentrates on the features more transferable across the datasets, the non-intensity features in this work, by disentangling those features and applying adversarial learning directly to them during the adaptation process. Example reconstruction results from the auto-encoding architecture as well as segmentation results in coronal view can be found in supplementary material.

\section{Conclusion}
We have presented a novel unsupervised domain adaptation method for small bowel segmentation. To increase the adaptability of a trained segmenter across different domains, we disentangle the feature representation into desired factors in an unsupervised way, and concentrate on adapting more transferable features among them. Finally, segmentation prediction is performed by aggregating the disentangled features. We evaluated our method using abdominal CT scans with and without oral contrast as the source and target domains, respectively. The experimental results showed clear improvement compared to other domain adaptation methods that are without the feature disentanglement. Considering the difficulty of labeling the small bowel, the obtained result is encouraging since the proposed method can adapt a model without using any target domain labels.

\subsubsection*{Acknowledgments.} We thank Dr. James Gulley for patient referral and for providing access to CT scans. This research was supported by the National Institutes of Health, Clinical Center.

%
%

%
%
%
%


%
\title{Unsupervised Domain Adaptation for\\Small Bowel Segmentation using\\Disentangled Representation:\\ Supplementary Material}
%
\titlerunning{Unsupervised Domain Adaptation for Small Bowel Segmentation}
%
\author{Seung Yeon Shin \and
	Sungwon Lee \and
	Ronald M. Summers}
\authorrunning{S.Y. Shin et al.}
%
\institute{Imaging Biomarkers and Computer-Aided Diagnosis Laboratory, Radiology and Imaging Sciences, Clinical Center, National Institutes of Health, Bethesda, MD, USA}

\maketitle              

\begin{figure}[h]
	\centering
	\begin{minipage}{1\textwidth}
		\subfloat{\includegraphics[width = 0.25\textwidth]{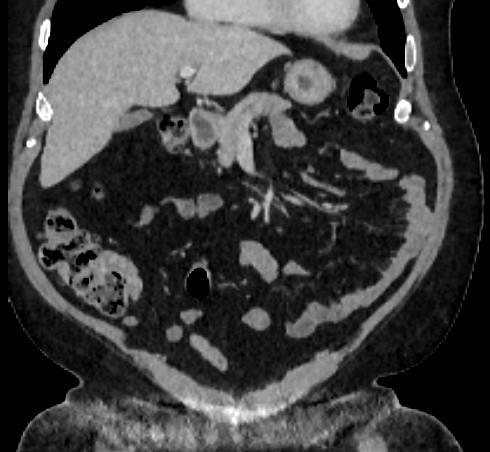}}
		\subfloat{\includegraphics[width = 0.25\textwidth]{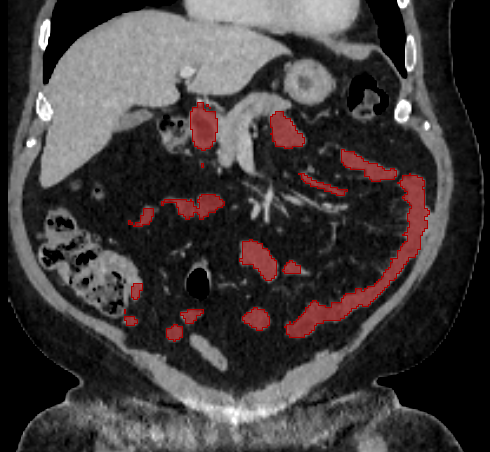}}
		\subfloat{\includegraphics[width = 0.25\textwidth]{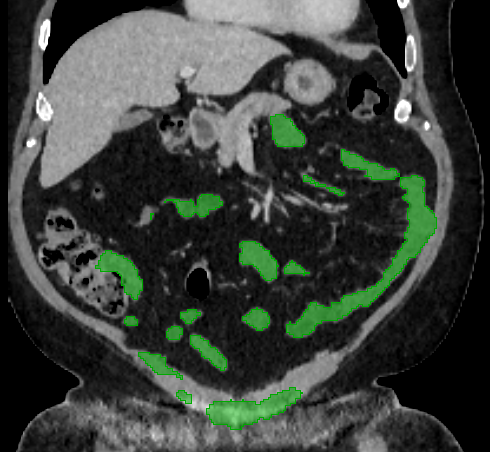}}
		\subfloat{\includegraphics[width = 0.25\textwidth]{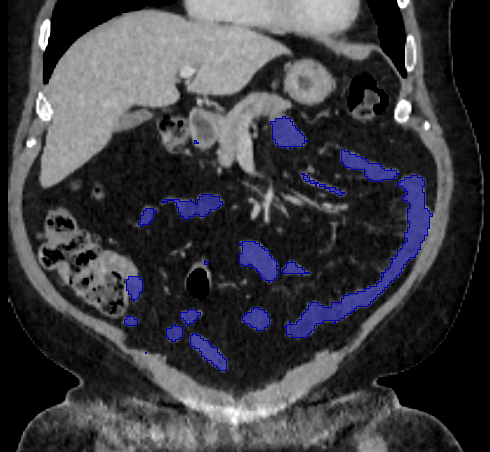}}
	\end{minipage}
	\begin{minipage}{1\textwidth}
		\subfloat{\includegraphics[width = 0.25\textwidth]{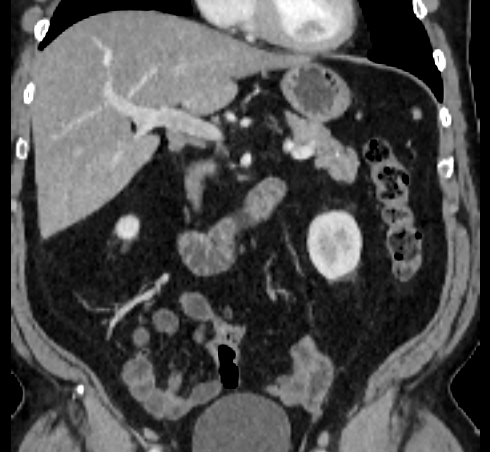}}
		\subfloat{\includegraphics[width = 0.25\textwidth]{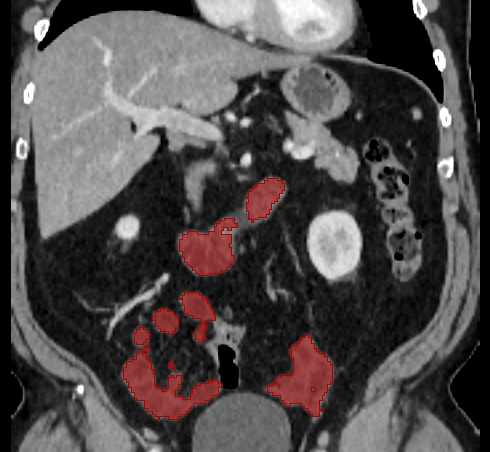}}
		\subfloat{\includegraphics[width = 0.25\textwidth]{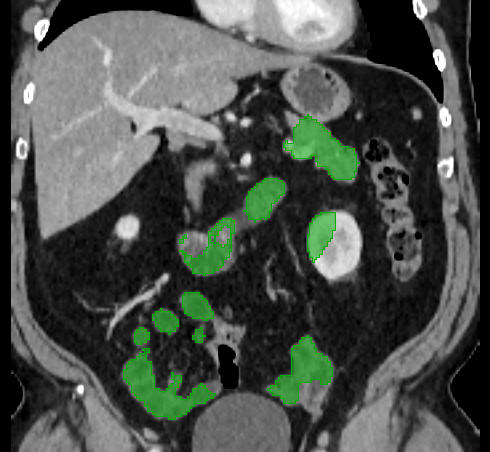}}
		\subfloat{\includegraphics[width = 0.25\textwidth]{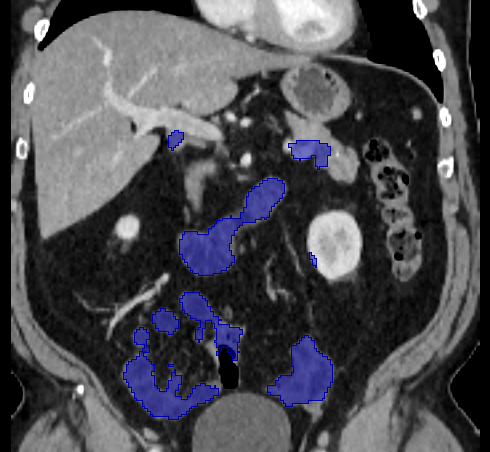}}
	\end{minipage}
	\begin{minipage}{1\textwidth}
		\subfloat{\includegraphics[width = 0.25\textwidth]{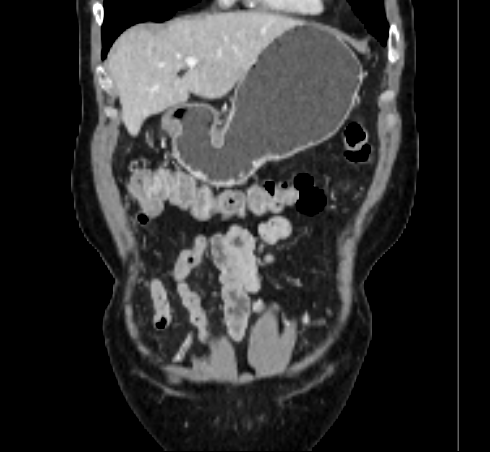}}
		\subfloat{\includegraphics[width = 0.25\textwidth]{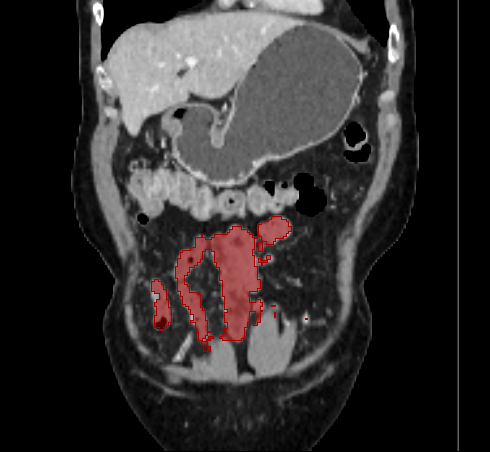}}
		\subfloat{\includegraphics[width = 0.25\textwidth]{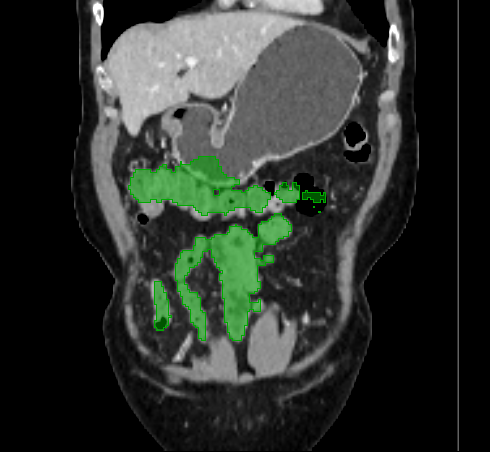}}
		\subfloat{\includegraphics[width = 0.25\textwidth]{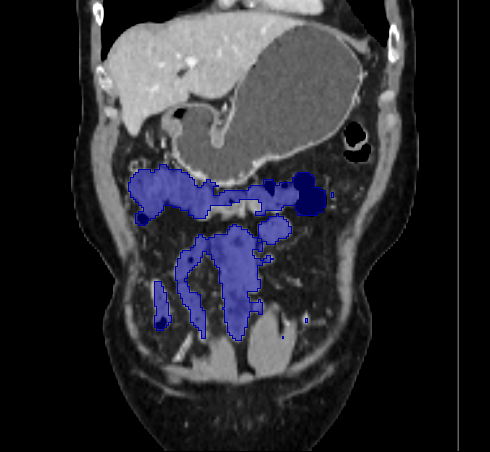}}
	\end{minipage}
	\caption{Example segmentation results in coronal view. Each row represents different scans. The columns, from left, represent an image slice of the input volume, ground-truth, result corresponding to `feat. \& out. level DA' in Table 1, and result of the proposed method. The proposed method produces a more accurate segmentation with fewer false positives, which are mostly in the large bowel, abdominal wall musculature, and kidney for the compared method. The last row is a failure case where false positives are observed for a part of the large bowel that has a similar appearance with the small bowel.}
	\label{fig:res_2d}
\end{figure}

\begin{figure}[t]
	\centering
	\subfloat[]{\includegraphics[width = 0.25\textwidth]{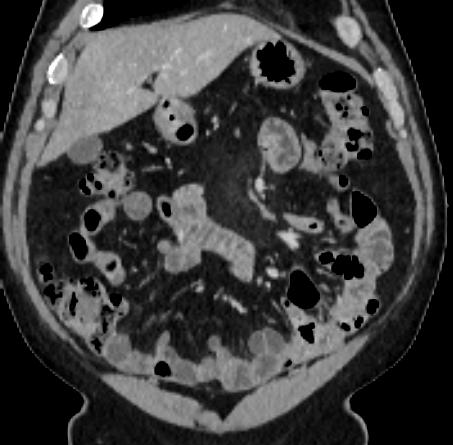}}
	\subfloat[]{\includegraphics[width = 0.25\textwidth]{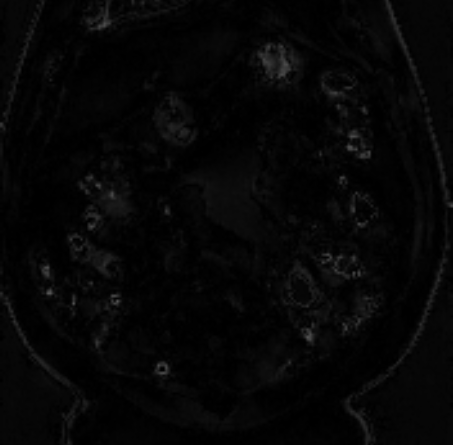}}
	\subfloat[]{\includegraphics[width = 0.25\textwidth]{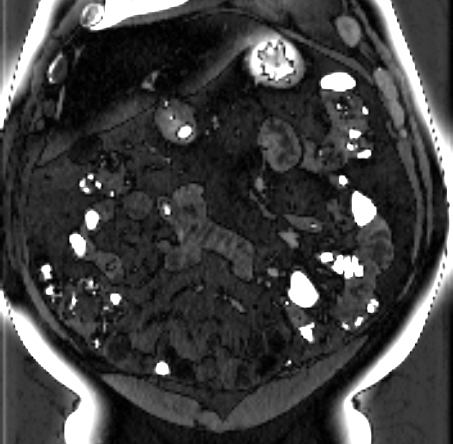}}
	\subfloat[]{\includegraphics[width = 0.25\textwidth]{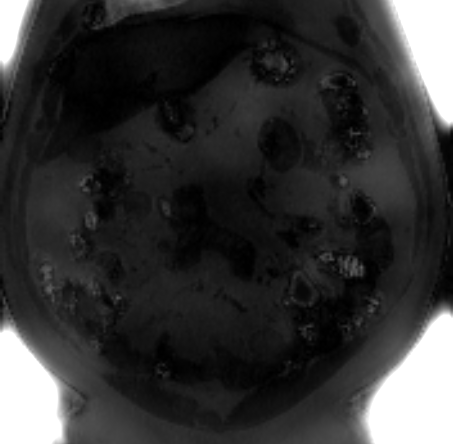}}
	\caption{Example reconstruction results. The images are: (a) an image slice of the input volume, (b) error map (difference between the input and the reconstructed volumes) when complete inputs $X$ and $\nabla X$ were given for the intensity and non-intensity encoders, respectively, (c) error map given only valid $X$, where a zero filled volume is used for $\nabla X$, and (d) error map given only valid $\nabla X$. The reconstruction from only the non-intensity features, (d), shows lower errors for the small bowel than the other region while (c) does not, meaning that the non-intensity features would have more information on the small bowel than the intensity features.}
	\label{fig:rec_comp}
\end{figure}

\end{document}